# Single-state semiquantum private comparison based on Bell states

Mao-Jie Geng, Ying Chen, Tian-Jie Xu, Tian-Yu Ye*

College of Information & Electronic Engineering, Zhejiang Gongshang University, Hangzhou 310018, P.R.China

E-mail：happyyty@aliyun.com(T.Y.Ye)

**Abstract:** In this paper, a novel semiquantum private comparison (SQPC) protocol based on single kind of Bell states is proposed, which allows two classical parties to judge the equality of their private inputs securely and correctly under the help of a semi-honest third party (TP) who possesses complete quantum capabilities. TP is allowed to misbehave on her own but cannot conspire with anyone else. Our protocol needs none of unitary operations, quantum entanglement swapping or the reordering operations. Moreover, our protocol only needs to prepare single kind of Bell states as initial quantum resource. Detailed security analysis turns out that our protocol is secure against various outside and participant attacks. Compared with most of the existing SQPC protocols based on Bell states, our protocol is more feasible in practice.

## 1 Introduction

Relying on the physical laws such as the quantum no-cloning theorem, the uncertainty principle *etc.*, quantum cryptography possesses unconditional security in theory. In 1984, Bennett and Brassard [1] proposed the world's first quantum key distribution (QKD) protocol by using the polarization of single photons, which is always named as the BB84 protocol later. Starting with the BB84 protocol, quantum cryptography has gained considerable developments in recent years. Researchers have proposed different kinds of quantum cryptography protocols suitable for different application scenarios, such as quantum key distribution (QKD) [2-6], quantum secret sharing (QSS) [7-10], quantum secure direct communication (QSDC) [11-16], quantum identity authentication (QIA) [17,18], and so on. QKD protocols can be used to establish a private sequence of key bits between two remote communicants through the transmission of quantum signals. QSS permits different participants to share a secret privately in the way that only all of them cooperate together can they reconstruct it. QSDC protocols can be used to directly transmit a secret message from one party to the other party. QIA protocols can be used for authenticating whether the user's identity is legal or not. In 2009, Yang and Wen [19] proposed the first quantum private comparison (QPC) protocol, which can compare the equality of private inputs from two different users under the condition that none of their private inputs will be leaked out. Since then, scholars have designed abundant QPC protocols with different quantum states, such as the ones with single particles [20,21], Bell states [22-24], GHZ states [25,26], cluster states [27], $\chi$-type entangled states [28], *etc*. The above QPC protocols require all users to have complete quantum capabilities, which may incur high costs in practice.

In 2007, Boyer *et al.* [29] proposed the first measure-resend semiquantum key distribution (SQKD) protocol by using the famous BB84 protocol, where it is not necessary for all users to have complete quantum capabilities. Subsequently, in 2009, Boyer *et al.* [30] further put forward the randomization-based SQKD protocol with single photons. At present, in the field of semiquantum cryptography, according to the two representative works in Refs.[29,30], classical users are widely considered to be limited within the following operations: (a) sending or reflecting the qubits without interference; (b) measuring qubits in the $Z$ basis (i.e., $\{|0\rangle, |1\rangle\}$); (c) preparing the fresh qubits in the $Z$ basis; and (d) reordering the qubits through different delay lines. Based on these settings, compared with the traditional quantum cryptography, semiquantum cryptography may effectively save quantum resource and quantum operations. The idea of semiquantumness has been applied into various branches of quantum cryptography so that the corresponding semiquantum cryptography branches have been established, such as SQKD [29-33], semiquantum secret sharing (SQSS) [34-37], semiquantum key agreement (SQKA) [38-41], semiquantum controlled secure direct communication (SQCSDC) [41] and semiquantum dialogue (SQD) [41,42], *etc*.

In 2016, Chou *et al.* [43] introduced the concept of semiquantumness into QPC and put forward the first semiquantum private comparison (SQPC) protocol by using Bell states and quantum entanglement swapping. In 2018, Thapliyal *et al.* [44] proposed a SQPC protocol based on Bell states, which needs to share a secret key in advance among different participants by using SQKD and SQKA; Ye *et al.* [45] designed a SQPC protocol with the measure-resend characteristics based on two-particle product states. In 2019, Yan *et al.* [39] constructed a randomization-based SQPC protocol based on Bell states without pre-shared keys; Lin *et al.* [46] proposed a SQPC protocol with single photons, which allows two classical participants to safely compare the equality of their private inputs under the help of an almost dishonest third party (TP); Yan *et al.* [47] put forward a SQPC scheme based on Greenberger-Horne-Zeilinger (GHZ) class states. In 2020, Jiang [48] put forward two SQPC protocols with Bell states, where the first protocol requires the classical users to measure the received particles but the second protocol doesn't have this requirement. In 2021, Tsai *et al.* [49] and Xie *et al.* [50] pointed out that the first SQPC protocol in Ref.[48] has security loopholes. In order to solve these problems, Ref.[49] makes TP share a secret key with each classical user in advance through SQKD, while Ref.[50] increases the quantum measurement capability for two classical users. In the same year, Yan *et al.* [51] put forward a SQPC protocol with three-particle GHZ-like states; Ye *et al.* [52] proposed an efficient circular SQPC protocol based on single-particle states without using a pre-shared key; Sun *et al.* [53] suggested a novel measure-resend SQPC scheme with pre-shared keys by using Bell states. Note that each of the SQPC protocols in Refs.[43,44,48-50,53] needs to employ four kinds of Bell states as initial quantum resource.

According to the above analysis, in this paper, in order to cut down the usage of initial quantum resource and quantum operations for SQPC based on Bell states, we propose a novel SQPC protocol based on single kind of Bell states, which utilizes the entanglement correlation of Bell states to skillfully compare the equality of two classical users' private inputs. Compared with most of the existing SQPC protocols with Bell states, our protocol is much easier to implement in practice, due to the following advantages: firstly, it only adopts one kind of Bell states as initial quantum resource; secondly, it doesn't need to employ quantum entanglement swapping or the reordering operations.

## 2  Protocol description

Suppose that both Alice and Bob only possess limited quantum capabilities; Alice has a private message $M_A$, while Bob has a private message $M_B$. Here, $M_A = \{M_A^1, M_A^2, \cdots, M_A^n\}$, $M_B = \{M_B^1, M_B^2, \cdots, M_B^n\}$, $M_A^i \in \{0,1\}$, $M_B^i \in \{0,1\}$, and $i = 0,1,\cdots,n$. Alice and Bob want to know whether $M_A$ and $M_B$ are equal or not under the help of TP who possesses full quantum capabilities. According to Ref.[54], here, TP is assumed to be semi-honest, which means that she is allowed to misbehave on her own but cannot conspire with anyone else. In order to securely compare the equality of $M_A$ and $M_B$ without disclosing their genuine contents to TP, Alice and Bob share a secret key $K_{AB} = \{K_{AB}^1, K_{AB}^2, \cdots, K_{AB}^n\}$ through a secure mediated SQKD protocol [31] in advance, where $K_{AB}^i \in \{0,1\}$ and $i = 0,1,\cdots,n$. The proposed SQPC protocol based on Bell states can be depicted as follows, whose flow chart is shown in Fig.1 for clarity.

Step 1: TP prepared $N = 16n$ Bell states all in the state of $|\psi^+\rangle = \frac{1}{\sqrt{2}}(|01\rangle + |10\rangle)$. Then, she divides the first particles and the second particles of the first $4n$ Bell states into sequences $T_1$ and $T_2$, respectively. Similarly, she makes the first particles and the second particles of the second $4n$ Bell states to form sequences $T_3$ and $T_4$, respectively, and lets the first particles and the second particles of the last $8n$ Bell states to compose sequences $T_5$ and $T_6$, respectively. Afterward, TP transmits the particles of sequence $T_1$ to Alice one by one. Note that after TP sends the first particle to Alice, she sends a particle only after receiving the previous one. In the meanwhile, TP transmits the particles of sequence $T_3$ to Bob one by one. Likewise, after TP sends the first particle to Bob, she sends a

particle only after receiving the previous one.

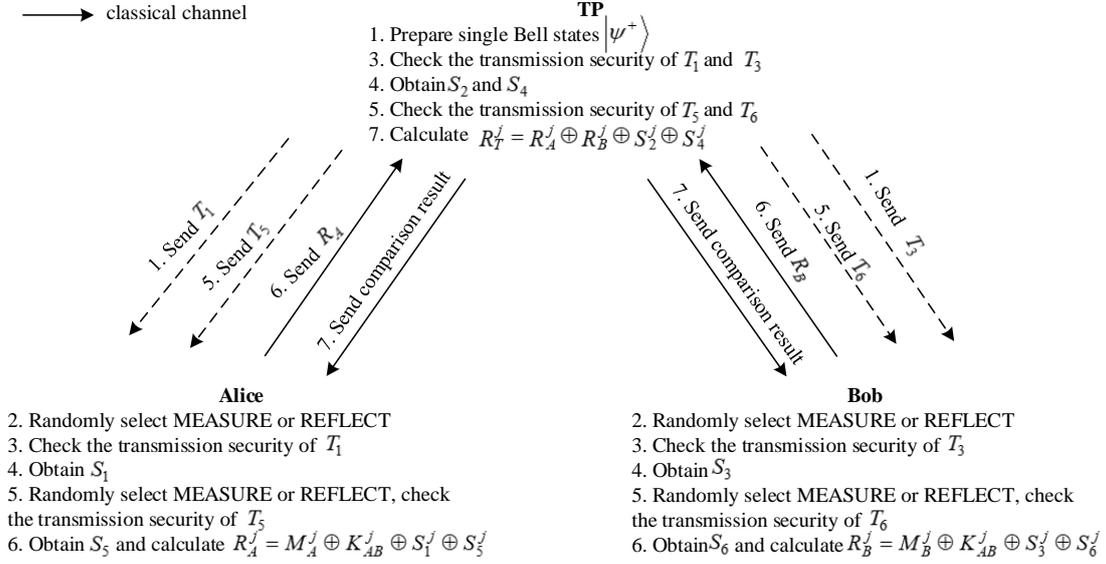

Fig. 1    The flowchart of the proposed single-state SQPC protocol

Step 2: After receiving each particle from TP, Alice randomly selects one of the following two operations: measuring the received particle with the $Z$ basis and resending a fresh one to TP in the same state to the measurement result (this is called as the MEASURE mode); and returning the received particle to TP directly (this is called as the REFLECT mode). Alice records the measurement results of corresponding particles in $T_1$ which she chose to MEASURE.

Similarly, after receiving each particle from TP, Bob randomly selects one of the following two operations, MEASURE and REFLECT. Bob also records the measurement results of corresponding particles in $T_3$ which he chose to MEASURE.

Step 3: TP and Alice cooperate together to check the transmission security of $T_1$. Alice tells TP the positions where she chose to REFLECT. TP performs different operations on the received particles according to Alice's choices, as illustrated in Table 1.

For checking the error rate on the $2n$ REFLECT particles in $T_1$, TP compares her Bell basis measurement results with her initial prepared states. Apparently, if there exists no Eve, TP's measurement results should always be $|\psi^+\rangle$.

For checking the error rate on the $2n$ MEASURE particles in $T_1$, TP randomly chooses $n$ ones among them, tells Alice the chosen positions and requires Alice to inform her of the corresponding measurement results on these $n$ chosen MEASURE particles in $T_1$. TP calculates the error rate by comparing her measurement results on these $n$ chosen MEASURE particles in $T_1$, her measurement results on the corresponding $n$ particles in $T_2$ and Alice's measurement results on these $n$ chosen MEASURE particles in $T_1$. Apparently, if there exists no Eve, these three kinds of measurement results should be perfectly correlated. For example, assuming that there exists no Eve, when Alice's measurement result on one chosen MEASURE particle in $T_1$ is $|0\rangle$ ($|1\rangle$),

TP's measurement result on this chosen MEASURE particle in $T_1$ should also be $|0\rangle$ ($|1\rangle$); and TP's measurement result on the corresponding particle in $T_2$ should be $|1\rangle$ ($|0\rangle$), since the corresponding initial Bell state prepared by TP is $|\psi^+\rangle$.

If either the error rate on the REFLECT particles or the error rate on the MEASURE particles in $T_1$ is unreasonably high, the protocol will be halted; otherwise, the protocol will be continued.

In the meanwhile, TP and Bob cooperate together to check the transmission security of $T_3$. Bob tells TP the positions where he chose to REFLECT. TP performs different operations on the received particles according to Bob's choices, as illustrated in Table 2.

For checking the error rate on the $2n$ REFLECT particles in $T_3$, TP compares her Bell basis measurement results with her initial prepared states.

For checking the error rate on the $2n$ MEASURE particles in $T_3$, TP randomly chooses $n$ ones among them, tells Bob the chosen positions and requires Bob to inform her of the corresponding measurement results on these $n$ chosen MEASURE particles in $T_3$. TP calculates the error rate by comparing her measurement results on these $n$ chosen MEASURE particles in $T_3$, her measurement results on the corresponding $n$ particles in $T_4$ and Bob's measurement results on these $n$ chosen MEASURE particles in $T_3$.

If either the error rate on the REFLECT particles or the error rate on the MEASURE particles in $T_3$ is unreasonably high, the protocol will be halted; otherwise, the protocol will be continued.

Table 1  Alice's actions on the particles in $T_1$ and TP's corresponding actions

| Case | Alice | TP |
|---|---|---|
| 1 | REFLECT | Action 1 |
| 2 | MEASURE | Action 2 |

Action 1: TP performs the Bell basis measurements on the REFLECT particles in $T_1$ and the corresponding particles in $T_2$;

Action 2: TP uses the $Z$ basis to measure the MEASURE particles in $T_1$ and the corresponding particles in $T_2$, respectively.

Table 2  Bob's actions on the particles in $T_3$ and TP's corresponding actions

| Case | Bob | TP |
|---|---|---|
| 1 | REFLECT | Action $1^\#$ |
| 2 | MEASURE | Action $2^\#$ |

Action $1^\#$: TP performs the Bell basis measurements on the REFLECT particles in $T_3$ and the corresponding particles in $T_4$;

Action $2^\#$: TP uses the $Z$ basis to measure the MEASURE particles in $T_3$ and the corresponding particles in $T_4$, respectively.

Step 4: TP drops out the $3n$ particles in $T_2$ used for security check in Step 3. TP's measurement results on the remaining $n$ particles in $T_2$ are represented as $S_2 = \{S_2^1, S_2^2, \cdots, S_2^n\}$. Alice's measurement results on these $n$ particles in $T_1$ are represented as $S_1 = \{S_1^1, S_1^2, \cdots, S_1^n\}$.

In the meanwhile, TP drops out the $3n$ particles in $T_4$ used for security check in Step 3. TP's measurement results on the remaining $n$ particles in $T_4$ are represented as $S_4 = \{S_4^1, S_4^2, \cdots, S_4^n\}$. Bob's measurement results on these $n$ particles in $T_3$ are represented as $S_3 = \{S_3^1, S_3^2, \cdots, S_3^n\}$.

Step 5: TP transmits the particles of sequence $T_5$ to Alice one by one. Note that after TP sends the first particle to Alice, she sends a particle only after receiving the previous one. After receiving each particle from TP, Alice randomly selects one of the two operations, MEASURE and REFLECT. Then, Alice tells TP the positions where she chose to REFLECT.

In the meanwhile, TP transmits the particles of sequence $T_6$ to Bob one by one. Note that after TP sends the first particle to Bob, she sends a particle only after receiving the previous one. After receiving each particle from TP, Bob randomly selects one of the two operations, MEASURE and REFLECT. Then, Bob tells TP the positions where he chose to REFLECT.

According to the choices of Alice and Bob, TP performs different operations on the received particles, as illustrated in Table 3. Case ①, Case ② and Case ③ are used for checking whether the transmissions of $T_5$ and $T_6$ are secure or not; and Case ④ is used for not only the security check of the transmissions of $T_5$ and $T_6$ but also privacy comparison.

For the $2n$ positions where Case ① happens, TP compares her Bell basis measurement results with her initial prepared states. Apparently, if there exists no Eve, TP's measurement results should always be $|\psi^+\rangle$.

For the $2n$ positions where Case ② happens, TP compares her $Z$ basis measurement results with Alice's $Z$ basis measurement results. Apparently, if there exists no Eve, TP's $Z$ basis measurement results should always be opposite to Alice's $Z$ basis measurement results.

For the $2n$ positions where Case ③ happens, TP compares her $Z$ basis measurement results with Bob's $Z$ basis measurement results. Apparently, if there exists no Eve, TP's $Z$ basis measurement results should always be opposite to Bob's $Z$ basis measurement results.

For the $2n$ positions where Case ④ happens, TP randomly chooses $n$ ones among them, and tells Alice and Bob the chosen positions. TP calculates the error rate by comparing her measurement results for the chosen positions with Alice and Bob's measurement results on them, respectively. Apparently, if there exists no Eve, TP's measurement results should be identical to Alice and Bob's measurement results, respectively.

If either of the error rates of Case ①, Case ②, Case ③ and Case ④ is unreasonably high, the protocol will be halted; otherwise, the protocol will be continued.

Table 3  Alice's actions on the particles in $T_5$, Bob's actions on the particles in $T_6$ and TP's corresponding actions

| Case | Alice | Bob | TP |
|---|---|---|---|
| ① | REFLECT | REFLECT | Action 1* |
| ② | MEASURE | REFLECT | Action 2* |
| ③ | REFLECT | MEASURE | Action 3* |
| ④ | MEASURE | MEASURE | Action 4* |

Action 1*: TP employs the Bell basis to measure the received particles in $T_5$ and the corresponding received particles in $T_6$;

Action 2*: TP measures the received particles in $T_6$ with the $Z$ basis, and requires Alice to inform her of the measurement results of corresponding particles in $T_5$;

Action 3*: TP measures the received particles in $T_5$ with the $Z$ basis, and requires Bob to inform her of the measurement results of corresponding particles in $T_6$;

Action 4*: TP measures the received particles in $T_5$ and $T_6$ with the $Z$ basis, respectively, and requires Alice and Bob to inform her of the measurement results of particles in $T_5$ and $T_6$, respectively.

Step 6: Alice drops out the $7n$ particles in $T_5$ used for security check in Step 5. In Case ④, Alice's measurement results on the remaining $n$ particles in $T_5$, which are not used for security check, are represented as $S_5 = \{S_5^1, S_5^2, \cdots, S_5^n\}$. Alice calculates $R_A^j = M_A^j \oplus K_{AB}^j \oplus S_1^j \oplus S_5^j$, where $\oplus$ is the bitwise XOR operation and $j = 1, 2, \cdots, n$. Then, Alice sends $R_A$ to TP, where $R_A = \{R_A^1, R_A^2, \cdots, R_A^n\}$.

In the meanwhile, Bob drops out the $7n$ particles in $T_6$ used for security check in Step 5. In Case ④, Bob's measurement results on the remaining $n$ particles in $T_6$, which are not used for security check, are represented as $S_6 = \{S_6^1, S_6^2, \cdots, S_6^n\}$. Bob calculates $R_B^j = M_B^j \oplus K_{AB}^j \oplus S_3^j \oplus S_6^j$, where $j = 1, 2, \cdots, n$. Then, Bob sends $R_B$ to TP, where $R_B = \{R_B^1, R_B^2, \cdots, R_B^n\}$.

Step 7: After receiving $R_A$ and $R_B$, TP computes $R_T^j = R_A^j \oplus R_B^j \oplus S_2^j \oplus S_4^j$, where $j = 1, 2, \cdots, n$. Once $R_T^j = 0$ is found, TP terminates the protocol, and tells Alice and Bob that $M_A$ is not identical to $M_B$. Otherwise, TP tells Alice and Bob that $M_A$ is same to $M_B$.

## 3  Correctness analysis

In the proposed SQPC protocol, Alice and Bob's secret messages are $M_A = \{M_A^1, M_A^2, \cdots, M_A^n\}$ and $M_B = \{M_B^1, M_B^2, \cdots, M_B^n\}$, respectively; Alice and Bob intent to judge the equality of $M_A$ and $M_B$ with the help of a semi-honest TP. Apparently, as the initial states prepared by TP are always in the state of $|\psi^+\rangle$, we have $S_1^j \oplus S_2^j = 1$, $S_3^j \oplus S_4^j = 1$ and $S_5^j \oplus S_6^j = 1$, where $j = 1, 2, \cdots, n$. Because $R_A^j = M_A^j \oplus K_{AB}^j \oplus S_1^j \oplus S_5^j$ and $R_B^j = M_B^j \oplus K_{AB}^j \oplus S_3^j \oplus S_6^j$, it can be obtained that

$$R_T^j = R_A^j \oplus R_B^j \oplus S_2^j \oplus S_4^j$$
$$= \left(M_A^j \oplus K_{AB}^j \oplus S_1^j \oplus S_5^j\right) \oplus \left(M_B^j \oplus K_{AB}^j \oplus S_3^j \oplus S_6^j\right) \oplus S_2^j \oplus S_4^j$$
$$= M_A^j \oplus M_B^j \oplus 1. \qquad (1)$$

According to Eq.(1), $R_T^j = 0$ indicates that $M_A^j \neq M_B^j$. Therefore, only when $R_T^j = 1$ stands for $j = 1, 2, \cdots, n$ can we have $M_A = M_B$.

## 4 Security analysis

### 4.1 Outside attack

In the proposed SQPC protocol, TP sends $T_1$ to Alice and $T_3$ to Bob firstly, and then transmits $T_5$ and $T_6$ to Alice and Bob, respectively. For clarity, we firstly analyze the security of transmission of $T_1$ or $T_3$, and then validate the security of transmissions of $T_5$ and $T_6$.

**Case 1: Eve attacks $T_1$ when it goes from TP to Alice or $T_3$ when it goes from TP to Bob**

In the proposed SQPC protocol, $T_1$, which is independent from $T_3$, essentially plays the same role to $T_3$. Thus, for simplicity, we only analyze the transmission security of $T_1$ from TP to Alice and back to TP.

(1) The intercept-resend attack

During the transmission of $T_1$ from TP to Alice and back to TP, Eve may try her best to obtain $S_1$ by launching the following attack: Eve may intercept the particles of $T_1$ from TP to Alice and send the fake particles she generated in the $Z$ basis beforehand to Alice. However, Eve will be inevitably detected due to the following two reasons: on one hand, the fake particles she prepared beforehand may be different from the genuine ones in $T_1$; and on the other hand, she doesn't know Alice's operations, which are random in fact. Concretely speaking, when TP sends one particle of $T_1$ to Alice, Eve intercepts it and sends the prepared fake one to Alice. Without loss of generality, assume that the fake one prepared by herself is in the state of $|0\rangle$. As a result, if Alice chooses to MEASURE, her measurement result will be $|0\rangle$. After Alice tells TP her operation, TP uses the $Z$ basis to measure the corresponding particle in $T_2$ and obtains the measurement result randomly in one of the two states $|0\rangle$ and $|1\rangle$. Hence, if Alice chooses to MEASURE, Eve will be detected with the probability of $\frac{1}{2}$. If Alice chooses to REFLECT, TP will receive the fake particle. After Alice tells TP her operation, TP measures the fake particle and the corresponding particle in $T_2$ with the Bell basis, and obtains the measurement result randomly in one of the four states $|\psi^+\rangle$, $|\psi^-\rangle$, $|\phi^+\rangle$ and $|\phi^-\rangle$, where $|\psi^-\rangle = \frac{1}{\sqrt{2}}(|01\rangle - |10\rangle)$ and $|\phi^\pm\rangle = \frac{1}{\sqrt{2}}(|00\rangle \pm |11\rangle)$. Hence, if Alice chooses to REFLECT, Eve will be detected with the probability of $\frac{3}{4}$. To sum up, when Eve launches this kind of attack on one particle of $T_1$, the probability that she will be discovered is $\frac{1}{2} \times \frac{1}{2} + \frac{1}{2} \times \frac{3}{4} = \frac{5}{8}$.

(2) The measure-resend attack

In order to obtain $S_1$, Eve may intercept the particles of $T_1$ sent from TP to Alice, measure them with the $Z$ basis and send the resulted states to Alice. However, this kind of attack from Eve will be inevitably discovered, since Alice's operations are random. Concretely speaking, after Eve's measurement, the particle Eve sends to Alice is randomly in one of the two states $|0\rangle$ and $|1\rangle$. Without loss of generality, assume that the particle after Eve's measurement is in the state of $|0\rangle$. If Alice chooses to MEASURE, after she tells TP her operation, TP uses the $Z$ basis to measure the corresponding particle in $T_2$ and obtains the measurement result $|1\rangle$. Hence, if Alice chooses to MEASURE, Eve will be detected with the probability of 0. If Alice chooses to REFLECT, after she tells TP her operation, TP measures the received particle from Alice and the corresponding particle in $T_2$ with the Bell basis, and obtains the measurement result randomly in one of the two states $|\psi^+\rangle$ and $|\psi^-\rangle$. Hence, if Alice chooses to REFLECT, Eve will be detected with the probability of $\frac{1}{2}$. To sum up, when Eve launches this kind of attack on one particle of $T_1$, the probability that she will be discovered is $\frac{1}{2} \times 0 + \frac{1}{2} \times \frac{1}{2} = \frac{1}{4}$.

(3) The entangle-measure attack

Eve may try to obtain something useful by entangling her auxiliary qubit with the transmitted qubit. Eve's entangle-measure attack on the particles of $T_1$, depicted as Fig.2, can be modeled with two unitary operations $\hat{E}$ and $\hat{F}$. Here, $\hat{E}$ attacks the qubit sent from TP to Alice, while $\hat{F}$ attacks the qubit returned from Alice to TP. Moreover, $\hat{E}$ and $\hat{F}$ share a common probe space with the initial state $|\varepsilon\rangle_E$. According to Refs.[29,30], the shared probe permits Eve to attack the returned particles depending on the knowledge obtained from $\hat{E}$; and any attack where Eve would make $\hat{F}$ depend on a measurement after performing $\hat{E}$ can be realized by $\hat{E}$ and $\hat{F}$ with controlled gates.

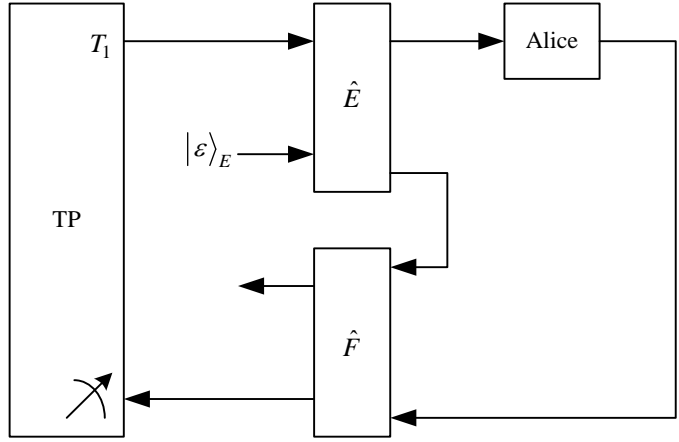

Fig.2 Eve's entangle-measure attack on the particles of $T_1$ with two unitaries $\hat{E}$ and $\hat{F}$

**Theorem 1.** *Suppose that Eve performs attack $(\hat{E},\hat{F})$ on the particle from TP to Alice and back to TP. For this attack inducing no error in Step 3, the final state of Eve's probe should be independent of not only Alice's operation but also TP and Alice's measurement results. As a result, Eve gets no information on the bits of $S_1$ and $S_2$.*

**Proof.** The effect of $\hat{E}$ on the qubits $|0\rangle$ and $|1\rangle$ can be expressed as

$$\hat{E}(|0\rangle|\varepsilon\rangle_E) = \alpha_{00}|0\rangle|\varepsilon_{00}\rangle + \alpha_{01}|1\rangle|\varepsilon_{01}\rangle, \tag{2}$$

$$\hat{E}(|1\rangle|\varepsilon\rangle_E) = \alpha_{10}|0\rangle|\varepsilon_{10}\rangle + \alpha_{11}|1\rangle|\varepsilon_{11}\rangle, \tag{3}$$

where $|\varepsilon_{00}\rangle$, $|\varepsilon_{01}\rangle$, $|\varepsilon_{10}\rangle$ and $|\varepsilon_{11}\rangle$ are Eve's probe states determined by $\hat{E}$, $|\alpha_{00}|^2 + |\alpha_{01}|^2 = 1$ and $|\alpha_{10}|^2 + |\alpha_{11}|^2 = 1$.

According to Stinespring dilation theorem, the global state of the composite system before Alice's operation is

$$\hat{E}(|\psi^+\rangle_{12}|\varepsilon\rangle_E) = \hat{E}\left[\frac{1}{\sqrt{2}}(|01\rangle + |10\rangle)_{12}|\varepsilon\rangle_E\right]$$

$$= \frac{1}{\sqrt{2}}\left[(\alpha_{00}|0\rangle_1|\varepsilon_{00}\rangle + \alpha_{01}|1\rangle_1|\varepsilon_{01}\rangle)|1\rangle_2 + (\alpha_{10}|0\rangle_1|\varepsilon_{10}\rangle + \alpha_{11}|1\rangle_1|\varepsilon_{11}\rangle)|0\rangle_2\right]$$

$$= \frac{1}{\sqrt{2}}\left[|0\rangle_1(\alpha_{00}|1\rangle_2|\varepsilon_{00}\rangle + \alpha_{10}|0\rangle_2|\varepsilon_{10}\rangle) + |1\rangle_1(\alpha_{01}|1\rangle_2|\varepsilon_{01}\rangle + \alpha_{11}|0\rangle_2|\varepsilon_{11}\rangle)\right], \tag{4}$$

where the subscripts 1 and 2 represent the particles from $T_1$ and $T_2$, respectively.

(i) Firstly, consider the case that Alice chooses to MEASURE. As a result, the state of the composite system is collapsed into $|0\rangle_1(\alpha_{00}|1\rangle_2|\varepsilon_{00}\rangle + \alpha_{10}|0\rangle_2|\varepsilon_{10}\rangle)$ when Alice's measurement result is $|0\rangle_1$ or $|1\rangle_1(\alpha_{01}|1\rangle_2|\varepsilon_{01}\rangle + \alpha_{11}|0\rangle_2|\varepsilon_{11}\rangle)$ when Alice's measurement result is $|1\rangle_1$.

Eve imposes $\hat{F}$ on the particle sent back to TP. In order that Eve's attacks on the MEASURE particle will not be discovered by TP and Alice in Step 3, the global state of the composite system should be

$$\hat{F}\left[|0\rangle_1 \left(\alpha_{00}|1\rangle_2|\varepsilon_{00}\rangle + \alpha_{10}|0\rangle_2|\varepsilon_{10}\rangle\right)\right] = |0\rangle_1|1\rangle_2|\varepsilon_0\rangle, \qquad (5)$$

when Alice's measurement result is $|0\rangle_1$; or

$$\hat{F}\left[|1\rangle_1 \left(\alpha_{01}|1\rangle_2|\varepsilon_{01}\rangle + \alpha_{11}|0\rangle_2|\varepsilon_{11}\rangle\right)\right] = |1\rangle_1|0\rangle_2|\varepsilon_1\rangle. \qquad (6)$$

when Alice's measurement result is $|1\rangle_1$.

(ii) Secondly, consider the case that Alice chooses to REFLECT. As a result, the state of the composite system is $\frac{1}{\sqrt{2}}\left[|0\rangle_1 \left(\alpha_{00}|1\rangle_2|\varepsilon_{00}\rangle + \alpha_{10}|0\rangle_2|\varepsilon_{10}\rangle\right) + |1\rangle_1 \left(\alpha_{01}|1\rangle_2|\varepsilon_{01}\rangle + \alpha_{11}|0\rangle_2|\varepsilon_{11}\rangle\right)\right]$.

Eve imposes $\hat{F}$ on the particle sent back to TP. According to Eq.(5) and Eq.(6), it has

$$\begin{aligned}
\hat{F}\left[\hat{E}\left(|\psi^+\rangle_{12}|\varepsilon\rangle_E\right)\right] &= \frac{1}{\sqrt{2}} \hat{F}\left[|0\rangle_1\left(\alpha_{00}|1\rangle_2|\varepsilon_{00}\rangle + \alpha_{10}|0\rangle_2|\varepsilon_{10}\rangle\right) + |1\rangle_1\left(\alpha_{01}|1\rangle_2|\varepsilon_{01}\rangle + \alpha_{11}|0\rangle_2|\varepsilon_{11}\rangle\right)\right] \\
&= \frac{1}{\sqrt{2}}\left(|0\rangle_1|1\rangle_2|\varepsilon_0\rangle + |1\rangle_1|0\rangle_2|\varepsilon_1\rangle\right) \\
&= \frac{1}{2}\left[\left(|\psi^+\rangle_{12} + |\psi^-\rangle_{12}\right)|\varepsilon_0\rangle + \left(|\psi^+\rangle_{12} - |\psi^-\rangle_{12}\right)|\varepsilon_1\rangle\right] \\
&= \frac{1}{2}\left[|\psi^+\rangle_{12}\left(|\varepsilon_0\rangle + |\varepsilon_1\rangle\right) + |\psi^-\rangle_{12}\left(|\varepsilon_0\rangle - |\varepsilon_1\rangle\right)\right]. \qquad (7)
\end{aligned}$$

In order that Eve's attacks on the REFLECT particle will not be discovered by TP and Alice in Step 3, the probability that TP's measurement result on a pair of particles in $T_1$ and $T_2$ is $|\psi^+\rangle_{12}$ should be 1. Thus, it should establish

$$|\varepsilon_0\rangle = |\varepsilon_1\rangle = |\varepsilon\rangle. \qquad (8)$$

Inserting Eq.(8) into Eq.(7) produces

$$\hat{F}\left[\hat{E}\left(|\psi^+\rangle_{12}|\varepsilon\rangle_E\right)\right] = |\psi^+\rangle_{12}|\varepsilon\rangle. \qquad (9)$$

(iii) Inserting Eq.(8) into Eq.(5) and Eq.(6) produces

$$\hat{F}\left[|0\rangle_1 \left(\alpha_{00}|1\rangle_2|\varepsilon_{00}\rangle + \alpha_{10}|0\rangle_2|\varepsilon_{10}\rangle\right)\right] = |0\rangle_1|1\rangle_2|\varepsilon\rangle, \qquad (10)$$

and

$$\hat{F}\left[|1\rangle_1 \left(\alpha_{01}|1\rangle_2|\varepsilon_{01}\rangle + \alpha_{11}|0\rangle_2|\varepsilon_{11}\rangle\right)\right] = |1\rangle_1|0\rangle_2|\varepsilon\rangle, \qquad (11)$$

respectively.

According to Eq.(9), Eq.(10) and Eq.(11), in order not to be detected by TP and Alice, the final state of Eve's probe should be independent from not only Alice's operation but also the TP and Alice's measurement results. As a result, Eve gets no information on the bits of $S_1$ and $S_2$.

（4）The Trojan horse attack

Because the particles of $T_1$ are transmitted forth and back between TP and Alice, two types of Trojan horse attack from Eve, i.e., the invisible photon eavesdropping attack [55] and the delay-photon Trojan horse attack [56,57], should be taken into account. In the light of Refs.[57,58], Alice can use a wavelength filter and a photon number splitter to resist these two types of Trojan horse attack from Eve, respectively.

**Case 2：Eve attacks $T_5$ and $T_6$ when they go from TP to Alice and Bob**

(1) The intercept-resend attack

In Step 5, TP transmits $T_5$ and $T_6$ to Alice and Bob, respectively. During these transmissions, Eve may try her best to obtain $S_5$ and $S_6$ by launching the following attack: Eve intercepts the two particles sent from TP to Alice and Bob, and sends two fake ones she prepared in the $Z$ basis beforehand to Alice and Bob, respectively. However, Eve will be undoubtedly discovered for two facts: firstly, the fake particles she prepared beforehand may be different from the genuine ones in $T_5$ and $T_6$; and secondly, Alice and Bob's operations are random. Concretely speaking, when TP sends one particle of $T_5$ to Alice and one particle of $T_6$ to Bob, Eve intercepts them and sends the prepared fake ones to Alice and Bob, respectively. Firstly, assume that the two fake particles prepared by Eve are in the state of $|0\rangle|0\rangle$. As a result, if both Alice and Bob choose to REFLECT, TP will receive the two fake particles $|0\rangle|0\rangle$. After Alice and Bob tell TP their operations, TP measures the two fake particles $|0\rangle|0\rangle$ with the Bell basis, and obtains the measurement result randomly in one of the two states $|\phi^+\rangle$ and $|\phi^-\rangle$. Hence, if both Alice and Bob choose to REFLECT, Eve will be detected with the probability of 1. If Alice chooses to MEASURE and Bob chooses to REFLECT, after Alice and Bob tell TP their operations, TP will measure the fake particle $|0\rangle$ reflected by Bob with the $Z$ basis and require Alice to announce her measurement result. Hence, if Alice chooses to MEASURE and Bob chooses to REFLECT, Eve will be detected with the probability of 1. Similarly, if Alice chooses to REFLECT and Bob chooses to MEASURE, Eve will be also detected with the probability of 1. If both Alice and Bob choose to MEASURE, after Alice and Bob tell TP their operations, TP will measure the received particles with the $Z$ basis and obtain the measurement result $|0\rangle|0\rangle$. So, if both Alice and Bob choose to MEASURE, Eve will be detected with the probability of 1. To sum up, when the two fake particles prepared by Eve are in the state of $|0\rangle|0\rangle$, Eve is detected with the probability of $\frac{1}{4}\times 1+\frac{1}{4}\times 1+\frac{1}{4}\times 1+\frac{1}{4}\times 1=1$. Secondly,

assume that the two fake particles prepared by Eve are in the state of $|1\rangle|1\rangle$. In this case, Eve is also detected with the probability of 1. Thirdly, assume that the two fake particles prepared by Eve are in the state of $|0\rangle|1\rangle$. As a result, if both Alice and Bob choose to REFLECT, TP will receive the two fake particles $|0\rangle|1\rangle$. After Alice and Bob tell TP their operations, TP measures the two fake particles $|0\rangle|1\rangle$ with the Bell basis, and obtains the measurement result randomly in one of the two states $|\psi^+\rangle$ and $|\psi^-\rangle$. Hence, if both Alice and Bob choose to REFLECT, Eve will be detected with the probability of $\frac{1}{2}$. If Alice chooses to MEASURE and Bob chooses to REFLECT, after Alice and Bob tell TP their operations, TP will measure the fake particle $|1\rangle$ reflected by Bob with the $Z$ basis and require Alice to announce her measurement result. Hence, if Alice chooses to MEASURE and Bob chooses to REFLECT, Eve will be detected with the probability of 0. Similarly, if Alice chooses to REFLECT and Bob chooses to MEASURE, Eve will be also detected with the probability of 0. If both Alice and Bob choose to MEASURE, after Alice and Bob tell TP their operations, TP will measure the received particles with the $Z$ basis and obtain the result $|0\rangle|1\rangle$. In this case, Eve will be detected with the probability of 0. To sum up, when the two fake particles prepared by TP are in the state of $|0\rangle|1\rangle$, Eve is detected with the probability of $\frac{1}{4} \times \frac{1}{2} + \frac{1}{4} \times 0 + \frac{1}{4} \times 0 + \frac{1}{4} \times 0 = \frac{1}{8}$. Fourthly, assume that the two fake particles prepared by Eve are in the state of $|1\rangle|0\rangle$. In this case, Eve is also detected with the probability of $\frac{1}{8}$.

(2) The measure-resend attack

In order to obtain $S_5$, Eve may intercept the particles of $T_5$ sent from TP to Alice, measure them with the $Z$ basis and send the resulted states to Alice. In the meanwhile, in order to obtain $S_6$, Eve may intercept the particles of $T_6$ sent from TP to Bob, measure them with the $Z$ basis and send the resulted states to Bob. However, this kind of attack from Eve will be detected undoubtedly as Alice and Bob's operations are random. Concretely speaking, after Eve's measurement, the Bell state prepared by TP is collapsed randomly into one of the two states $|01\rangle$ and $|10\rangle$. Without loss of generality, assume that the Bell state from TP after Eve's measurement is collapsed into $|01\rangle$. If both Alice and Bob choose to REFLECT, after Alice and Bob tell TP their operations, TP will measure the received particles $|01\rangle$ with the Bell basis and obtain $|\psi^+\rangle$ or $|\psi^-\rangle$ with equal probability. As a result, if both Alice and Bob choose to REFLECT, Eve will be detected with the probability of $\frac{1}{2}$. If Alice chooses to MEASURE and Bob chooses to REFLECT, after Alice and Bob tell TP their operations, TP will measure the particle $|1\rangle$ reflected by Bob with the $Z$ basis and require Alice to announce her measurement result. Hence, if Alice chooses to MEASURE and Bob chooses to REFLECT, Eve will be detected with the probability of 0. Similarly, if Alice chooses to REFLECT and Bob chooses to MEASURE, Eve will be also discovered with the probability of 0. If both Alice and Bob choose to MEASURE, after Alice and Bob tell TP their operations, TP will measure the received particles with the $Z$ basis. Therefore, if both Alice and Bob choose to MEASURE, Eve will be detected with the probability of 0. To sum up, when Eve launches this kind of attack on one particle of $T_5$ and one particle of $T_6$, the probability that she will be discovered is $\frac{1}{4} \times \frac{1}{2} + \frac{1}{4} \times 0 + \frac{1}{4} \times 0 + \frac{1}{4} \times 0 = \frac{1}{8}$.

(3) The entangle-measure attack

Eve's entangle-measure attack on the particles of $T_5$ and $T_6$, described as Fig.3, can be modeled as two unitaries: $U_E$ attacking particles from TP to Alice and Bob and $U_F$ attacking particles back from Alice and Bob to TP, where $U_E$ and $U_F$ share a common probe space with initial state $|\xi\rangle_E$.

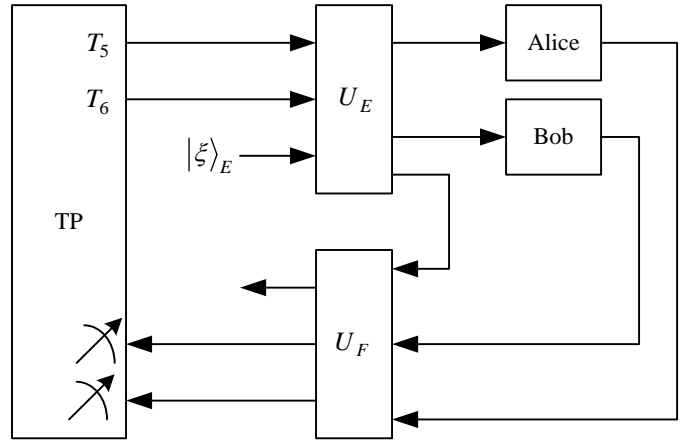

Fig.3  Eve's entangle-measure attack on the particles of $T_5$ and $T_6$ with two unitaries $U_E$ and $U_F$

**Theorem 2.** *Suppose that Eve performs attack $(U_E, U_F)$ on the particles from TP to Alice and Bob and back to TP. For this attack inducing no error in Step 5, the final state of Eve's probe should be independent of not only Alice and Bob's operations but also their measurement results. As a result, Eve gets no information on the bits of $S_5$ and $S_6$.*

**Proof.** The effect of $U_E$ on the qubits $|0\rangle$ and $|1\rangle$ can be expressed as

$$U_E(|0\rangle|\xi\rangle_E) = \beta_{00}|0\rangle|\xi_{00}\rangle + \beta_{01}|1\rangle|\xi_{01}\rangle, \tag{12}$$

$$U_E(|1\rangle|\xi\rangle_E) = \beta_{10}|0\rangle|\xi_{10}\rangle + \beta_{11}|1\rangle|\xi_{11}\rangle, \tag{13}$$

where $|\xi_{00}\rangle, |\xi_{01}\rangle, |\xi_{10}\rangle$ and $|\xi_{11}\rangle$ are Eve's probe states determined by $U_E$, $|\beta_{00}|^2 + |\beta_{01}|^2 = 1$ and $|\beta_{10}|^2 + |\beta_{11}|^2 = 1$.

According to Stinespring dilation theorem, the global state of the composite system before Alice and Bob's operations is

$$
\begin{aligned}
U_E\left(|\psi^+\rangle_{56}|\xi\rangle_E\right) &= U_E\left[\frac{1}{\sqrt{2}}(|01\rangle + |10\rangle)_{56}|\xi\rangle_E\right] \\
&= \frac{1}{\sqrt{2}}\Big[(\beta_{00}|0\rangle_5|\xi_{00}\rangle + \beta_{01}|1\rangle_5|\xi_{01}\rangle)(\beta_{10}|0\rangle_6|\xi_{10}\rangle + \beta_{11}|1\rangle_6|\xi_{11}\rangle) \\
&\quad + (\beta_{10}|0\rangle_5|\xi_{10}\rangle + \beta_{11}|1\rangle_5|\xi_{11}\rangle)(\beta_{00}|0\rangle_6|\xi_{00}\rangle + \beta_{01}|1\rangle_6|\xi_{01}\rangle)\Big] \\
&= \frac{1}{\sqrt{2}}\Big[|0\rangle_5|0\rangle_6(\beta_{00}\beta_{10}|\xi_{00}\rangle|\xi_{10}\rangle + \beta_{10}\beta_{00}|\xi_{10}\rangle|\xi_{00}\rangle) \\
&\quad + |0\rangle_5|1\rangle_6(\beta_{00}\beta_{11}|\xi_{00}\rangle|\xi_{11}\rangle + \beta_{10}\beta_{01}|\xi_{10}\rangle|\xi_{01}\rangle) \\
&\quad + |1\rangle_5|0\rangle_6(\beta_{01}\beta_{10}|\xi_{01}\rangle|\xi_{10}\rangle + \beta_{11}\beta_{00}|\xi_{11}\rangle|\xi_{00}\rangle) \\
&\quad + |1\rangle_5|1\rangle_6(\beta_{01}\beta_{11}|\xi_{01}\rangle|\xi_{11}\rangle + \beta_{11}\beta_{01}|\xi_{11}\rangle|\xi_{01}\rangle)\Big] \\
&= |0\rangle_5|0\rangle_6|\vartheta_{00}\rangle + |0\rangle_5|1\rangle_6|\vartheta_{01}\rangle + |1\rangle_5|0\rangle_6|\vartheta_{10}\rangle + |1\rangle_5|1\rangle_6|\vartheta_{11}\rangle,
\end{aligned}
\tag{14}
$$

where the subscripts 5 and 6 represent the particles from $T_5$ and $T_6$, respectively, and

$$|\vartheta_{00}\rangle = \frac{1}{\sqrt{2}}(\beta_{00}\beta_{10}|\xi_{00}\rangle|\xi_{10}\rangle + \beta_{10}\beta_{00}|\xi_{10}\rangle|\xi_{00}\rangle), \tag{15}$$

$$|\vartheta_{01}\rangle = \frac{1}{\sqrt{2}}(\beta_{00}\beta_{11}|\xi_{00}\rangle|\xi_{11}\rangle + \beta_{10}\beta_{01}|\xi_{10}\rangle|\xi_{01}\rangle), \tag{16}$$

$$|\vartheta_{10}\rangle = \frac{1}{\sqrt{2}}(\beta_{01}\beta_{10}|\xi_{01}\rangle|\xi_{10}\rangle + \beta_{11}\beta_{00}|\xi_{11}\rangle|\xi_{00}\rangle), \tag{17}$$

$$|\vartheta_{11}\rangle = \frac{1}{\sqrt{2}}(\beta_{01}\beta_{11}|\xi_{01}\rangle|\xi_{11}\rangle + \beta_{11}\beta_{01}|\xi_{11}\rangle|\xi_{01}\rangle). \tag{18}$$

When Alice and Bob receive the particles from TP, they choose either to MEASURE or to REFLECT. After that, Eve performs $U_F$ on the particles sent back to TP.

(i) Consider the case that both Alice and Bob choose to MEASURE. As a result, the state of the composite system is collapsed into $|x\rangle_5|y\rangle_6|\vartheta_{xy}\rangle$, where $x, y \in \{0,1\}$. For Eve not being detectable in Step 5, $U_F$ should satisfy

$$U_F(|x\rangle_5|y\rangle_6|\vartheta_{xy}\rangle) = |x\rangle_5|y\rangle_6|\gamma_{xy}\rangle, \tag{19}$$

which means that $U_F$ cannot alter the state of particles from Alice and Bob. Otherwise, Eve is discovered with a non-zero probability.

(ii) Consider the case that Alice chooses to MEASURE and Bob chooses to REFLECT. As a result, the state of the composite system is collapsed into $|0\rangle_5|0\rangle_6|\vartheta_{00}\rangle+|0\rangle_5|1\rangle_6|\vartheta_{01}\rangle$ when Alice's measurement result is $|0\rangle_5$ or $|1\rangle_5|0\rangle_6|\vartheta_{10}\rangle+|1\rangle_5|1\rangle_6|\vartheta_{11}\rangle$ when Alice's measurement result is $|1\rangle_5$.

Assume that Alice's measurement result is $|0\rangle_5$. After Eve imposes $U_F$ on the particles sent back to TP, due to Eq.(19), the state of the composite system is evolved into

$$U_F\left(|0\rangle_5|0\rangle_6|\vartheta_{00}\rangle+|0\rangle_5|1\rangle_6|\vartheta_{01}\rangle\right)=|0\rangle_5|0\rangle_6|\gamma_{00}\rangle+|0\rangle_5|1\rangle_6|\gamma_{01}\rangle. \tag{20}$$

For Eve not being detectable in Step 5, the probability that TP's measurement result on the particle reflected by Bob is $|0\rangle_6$ should be 0. Hence, it has

$$|\gamma_{00}\rangle=0. \tag{21}$$

On the other hand, assume that Alice's measurement result is $|1\rangle_5$. After Eve imposes $U_F$ on the particles sent back to TP, due to Eq.(19), the state of the composite system is evolved into

$$U_F\left(|1\rangle_5|0\rangle_6|\vartheta_{10}\rangle+|1\rangle_5|1\rangle_6|\vartheta_{11}\rangle\right)=|1\rangle_5|0\rangle_6|\gamma_{10}\rangle+|1\rangle_5|1\rangle_6|\gamma_{11}\rangle. \tag{22}$$

For Eve not being detectable in Step 5, the probability that TP's measurement result on the particle reflected by Bob is $|1\rangle_6$ should be 0. Hence, it has

$$|\gamma_{11}\rangle=0. \tag{23}$$

(iii) Consider the case that Alice chooses to REFLECT and Bob chooses to MEASURE. As a result, the state of the composite system is collapsed into $|0\rangle_5|0\rangle_6|\vartheta_{00}\rangle+|1\rangle_5|0\rangle_6|\vartheta_{10}\rangle$ when Bob's measurement result is $|0\rangle_6$ or $|0\rangle_5|1\rangle_6|\vartheta_{01}\rangle+|1\rangle_5|1\rangle_6|\vartheta_{11}\rangle$ when Bob's measurement result is $|1\rangle_6$.

Assume that Bob's measurement result is $|0\rangle_6$. After Eve imposes $U_F$ on the particles sent back to TP, due to Eq.(19), the state of the composite system is evolved into

$$U_F\left(|0\rangle_5|0\rangle_6|\vartheta_{00}\rangle+|1\rangle_5|0\rangle_6|\vartheta_{10}\rangle\right)=|0\rangle_5|0\rangle_6|\gamma_{00}\rangle+|1\rangle_5|0\rangle_6|\gamma_{10}\rangle. \tag{24}$$

For Eve not being detectable in Step 5, the probability that TP's measurement result on the particle reflected by Alice is $|0\rangle_5$ should be 0. This automatically stands after Eq.(21) is inserted into Eq.(24).

On the other hand, assume that Bob's measurement result is $|1\rangle_6$. After Eve imposes $U_F$ on the particles sent back to TP, due to Eq.(19), the state of the composite system is evolved into

$$U_F\left(|0\rangle_5|1\rangle_6|\vartheta_{01}\rangle+|1\rangle_5|1\rangle_6|\vartheta_{11}\rangle\right)=|0\rangle_5|1\rangle_6|\gamma_{01}\rangle+|1\rangle_5|1\rangle_6|\gamma_{11}\rangle. \tag{25}$$

For Eve not being detectable in Step 5, the probability that TP's measurement result on the particle reflected by Alice is $|1\rangle_5$ should be 0. This automatically stands after Eq.(23) is inserted into Eq.(25).

(iv) Consider the case that both Alice and Bob choose to REFLECT. As a result, the state of the composite system is $|0\rangle_5|0\rangle_6|\vartheta_{00}\rangle+|0\rangle_5|1\rangle_6|\vartheta_{01}\rangle+|1\rangle_5|0\rangle_6|\vartheta_{10}\rangle+|1\rangle_5|1\rangle_6|\vartheta_{11}\rangle$. After Eve imposes $U_F$ on the particles sent back to TP, due to Eq.(19), the state of the composite system is evolved into

$$U_F\left(|0\rangle_5|0\rangle_6|\vartheta_{00}\rangle+|0\rangle_5|1\rangle_6|\vartheta_{01}\rangle+|1\rangle_5|0\rangle_6|\vartheta_{10}\rangle+|1\rangle_5|1\rangle_6|\vartheta_{11}\rangle\right)=$$

$$|0\rangle_5|0\rangle_6|\gamma_{00}\rangle+|0\rangle_5|1\rangle_6|\gamma_{01}\rangle+|1\rangle_5|0\rangle_6|\gamma_{10}\rangle+|1\rangle_5|1\rangle_6|\gamma_{11}\rangle. \quad (26)$$

After Eq.(21) and Eq.(23) are inserted into Eq.(26), it can be obtained that

$$U_F\left(|0\rangle_5|0\rangle_6|\vartheta_{00}\rangle+|0\rangle_5|1\rangle_6|\vartheta_{01}\rangle+|1\rangle_5|0\rangle_6|\vartheta_{10}\rangle+|1\rangle_5|1\rangle_6|\vartheta_{11}\rangle\right)$$

$$=|0\rangle_5|1\rangle_6|\gamma_{01}\rangle+|1\rangle_5|0\rangle_6|\gamma_{10}\rangle$$

$$=\frac{1}{\sqrt{2}}\left(|\psi^+\rangle_{56}+|\psi^-\rangle_{56}\right)|\gamma_{01}\rangle+\frac{1}{\sqrt{2}}\left(|\psi^+\rangle_{56}-|\psi^-\rangle_{56}\right)|\gamma_{10}\rangle$$

$$=\frac{1}{\sqrt{2}}|\psi^+\rangle_{56}\left(|\gamma_{01}\rangle+|\gamma_{10}\rangle\right)+\frac{1}{\sqrt{2}}|\psi^-\rangle_{56}\left(|\gamma_{01}\rangle-|\gamma_{10}\rangle\right). \quad (27)$$

For Eve not being detectable in Step 5, the probability that TP's measurement result on the particles reflected by Alice and Bob is $|\psi^+\rangle_{56}$ should be 1. Therefore, according to Eq.(27), it has

$$|\gamma_{01}\rangle=|\gamma_{10}\rangle=|\gamma\rangle. \quad (28)$$

Inserting Eq.(28) into Eq.(27) produces

$$U_F\left(|0\rangle_5|0\rangle_6|\vartheta_{00}\rangle+|0\rangle_5|1\rangle_6|\vartheta_{01}\rangle+|1\rangle_5|0\rangle_6|\vartheta_{10}\rangle+|1\rangle_5|1\rangle_6|\vartheta_{11}\rangle\right)=\sqrt{2}|\psi^+\rangle_{56}|\gamma\rangle. \quad (29)$$

(v) Applying Eq.(28) into Eq.(19) produces

$$U_F\left(|0\rangle_5|1\rangle_6|\vartheta_{01}\rangle\right)=|0\rangle_5|1\rangle_6|\gamma\rangle, \quad (30)$$

$$U_F\left(|1\rangle_5|0\rangle_6|\vartheta_{10}\rangle\right)=|1\rangle_5|0\rangle_6|\gamma\rangle. \quad (31)$$

Applying Eq.(21) and Eq.(28) into Eq.(20) produces

$$U_F\left(|0\rangle_5|0\rangle_6|\vartheta_{00}\rangle+|0\rangle_5|1\rangle_6|\vartheta_{01}\rangle\right)=|0\rangle_5|1\rangle_6|\gamma\rangle. \quad (32)$$

Applying Eq.(23) and Eq.(28) into Eq.(22) produces

$$U_F\left(|1\rangle_5|0\rangle_6|\vartheta_{10}\rangle+|1\rangle_5|1\rangle_6|\vartheta_{11}\rangle\right)=|1\rangle_5|0\rangle_6|\gamma\rangle. \quad (33)$$

Applying Eq.(21) and Eq.(28) into Eq.(24) produces

$$U_F\left(|0\rangle_5|0\rangle_6|\vartheta_{00}\rangle+|1\rangle_5|0\rangle_6|\vartheta_{10}\rangle\right)=|1\rangle_5|0\rangle_6|\gamma\rangle. \tag{34}$$

Applying Eq.(23) and Eq.(28) into Eq.(25) produces

$$U_F\left(|0\rangle_5|1\rangle_6|\vartheta_{01}\rangle+|1\rangle_5|1\rangle_6|\vartheta_{11}\rangle\right)=|0\rangle_5|1\rangle_6|\gamma\rangle. \tag{35}$$

According to Eqs.(29-35), it can be concluded that for Eve not inducing an error in Step 5, the final state of Eve's probe should be independent of not only Alice and Bob's operations but also their measurement results. As a result, Eve gets no information on the bits of $S_5$ and $S_6$.

Secondly, we consider the entangle-measure attack from Eve that she only performs $U_E$ on the particles of $T_5$ and $T_6$ sent out from TP.

**Lemma 1.** *Suppose that Eve only performs $U_E$ on the particles of $T_5$ and $T_6$ sent out from TP. For this attack inducing no error in Step 5, the final state of Eve's probe should be independent of not only Alice and Bob's operations but also their measurement results. As a result, Eve gets no information on the bits of $S_5$ and $S_6$.*

**Proof.** The global state of the composite system before Alice and Bob's operations can be depicted as Eq.(14). When Alice and Bob receive the particles from TP, they choose either to MEASURE or to REFLECT.

(i) Consider the case that both Alice and Bob choose to MEASURE. As a result, the state of the composite system is collapsed into $|x\rangle_5|y\rangle_6|\vartheta_{xy}\rangle$, where $x,y \in \{0,1\}$. After that, Eve does nothing on the particles sent back to TP. Hence, Eve cannot be detected in Step 5.

(ii) Consider the case that Alice chooses to MEASURE and Bob chooses to REFLECT. As a result, the state of the composite system is collapsed into $|0\rangle_5|0\rangle_6|\vartheta_{00}\rangle+|0\rangle_5|1\rangle_6|\vartheta_{01}\rangle$ when Alice's measurement result is $|0\rangle_5$ or $|1\rangle_5|0\rangle_6|\vartheta_{10}\rangle+|1\rangle_5|1\rangle_6|\vartheta_{11}\rangle$ when Alice's measurement result is $|1\rangle_5$.

Assume that Alice's measurement result is $|0\rangle_5$. Eve does nothing on the particles sent back to TP. For Eve not being detectable in Step 5, the probability that TP's measurement result on the particle reflected by Bob is $|0\rangle_6$ should be 0. Hence, it has

$$|\vartheta_{00}\rangle=0. \tag{36}$$

On the other hand, assume that Alice's measurement result is $|1\rangle_5$. Eve does nothing on the particles sent back to TP. For Eve not being detectable in Step 5, the probability that TP's measurement result on the particle reflected by Bob is $|1\rangle_6$ should be 0. Hence, it has

$$|\vartheta_{11}\rangle=0. \tag{37}$$

(iii) Consider the case that Alice chooses to REFLECT and Bob chooses to MEASURE. As a

result, the state of the composite system is collapsed into $|0\rangle_5|0\rangle_6|\vartheta_{00}\rangle+|1\rangle_5|0\rangle_6|\vartheta_{10}\rangle$ when Bob's measurement result is $|0\rangle_6$ or $|0\rangle_5|1\rangle_6|\vartheta_{01}\rangle+|1\rangle_5|1\rangle_6|\vartheta_{11}\rangle$ when Bob's measurement result is $|1\rangle_6$.

Assume that Bob's measurement result is $|0\rangle_6$. Eve does nothing on the particles sent back to TP. For Eve not being detectable in Step 5, the probability that TP's measurement result on the particle reflected by Alice is $|0\rangle_5$ should be 0. This is automatically satisfied, since Eq.(36) stands.

On the other hand, assume that Bob's measurement result is $|1\rangle_6$. Eve does nothing on the particles sent back to TP. For Eve not being detectable in Step 5, the probability that TP's measurement result on the particle reflected by Alice is $|1\rangle_5$ should be 0. This is automatically satisfied, since Eq.(37) stands.

(iv) Consider the case that both Alice and Bob choose to REFLECT. As a result, the state of the composite system is $|0\rangle_5|0\rangle_6|\vartheta_{00}\rangle+|0\rangle_5|1\rangle_6|\vartheta_{01}\rangle+|1\rangle_5|0\rangle_6|\vartheta_{10}\rangle+|1\rangle_5|1\rangle_6|\vartheta_{11}\rangle$. Because of Eq.(36) and Eq.(37), the state of the composite system can be expressed as

$$|0\rangle_5|0\rangle_6|\vartheta_{00}\rangle+|0\rangle_5|1\rangle_6|\vartheta_{01}\rangle+|1\rangle_5|0\rangle_6|\vartheta_{10}\rangle+|1\rangle_5|1\rangle_6|\vartheta_{11}\rangle=|0\rangle_5|1\rangle_6|\vartheta_{01}\rangle+|1\rangle_5|0\rangle_6|\vartheta_{10}\rangle$$

$$=\frac{1}{\sqrt{2}}\left(|\psi^+\rangle_{56}+|\psi^-\rangle_{56}\right)|\vartheta_{01}\rangle+\frac{1}{\sqrt{2}}\left(|\psi^+\rangle_{56}-|\psi^-\rangle_{56}\right)|\vartheta_{10}\rangle$$

$$=\frac{1}{\sqrt{2}}|\psi^+\rangle_{56}\left(|\vartheta_{01}\rangle+|\vartheta_{10}\rangle\right)+\frac{1}{\sqrt{2}}|\psi^-\rangle_{56}\left(|\vartheta_{01}\rangle-|\vartheta_{10}\rangle\right). \quad (38)$$

Eve does nothing on the particles sent back to TP. For Eve not being detectable in Step 5, the probability that TP's measurement result on the particles reflected by Alice and Bob is $|\psi^+\rangle_{56}$ should be 1. Therefore, it can be derived from Eq.(38) that

$$|\vartheta_{01}\rangle=|\vartheta_{10}\rangle=|\vartheta\rangle. \quad (39)$$

Inserting Eq.(39) into Eq.(38) produces

$$|0\rangle_5|0\rangle_6|\vartheta_{00}\rangle+|0\rangle_5|1\rangle_6|\vartheta_{01}\rangle+|1\rangle_5|0\rangle_6|\vartheta_{10}\rangle+|1\rangle_5|1\rangle_6|\vartheta_{11}\rangle=\sqrt{2}|\psi^+\rangle_{56}|\vartheta\rangle. \quad (40)$$

It can be concluded from the above analysis: considering the entangle-measure attack from Eve that she only performs $U_E$ on the particles of $T_5$ and $T_6$ sent out from TP, for Eve not inducing an error in Step 5, the final state of Eve's probe should be independent of not only Alice and Bob's operations but also their measurement results. As a result, Eve gets no information on the bits of $S_5$ and $S_6$.

Thirdly, we consider the entangle-measure attack from Eve that she only performs $U_F$ on the particles of $T_5$ and $T_6$ sent back to TP.

**Lemma 2.** *Suppose that Eve only performs $U_F$ on the particles of $T_5$ and $T_6$ sent back to TP. For this attack inducing no error in Step 5, the final state of Eve's probe should be independent of not only Alice and Bob's operations but also their measurement results. As a result, Eve gets no information on the bits of $S_5$ and $S_6$.*

**Proof.** Eve does nothing on the particles of $T_5$ and $T_6$ sent out from TP. As a result, the global state of the composite system before Alice and Bob's operations can be depicted as $|\psi^+\rangle_{56}|\xi\rangle_E$. When Alice and Bob receive the particles sent out from TP, they choose either to MEASURE or to REFLECT.

(i) Consider the case that both Alice and Bob choose to MEASURE. As a result, the state of the composite system is randomly collapsed into $|t\rangle_5|r\rangle_6|\xi\rangle_E$, where $t, r \in \{0,1\}$ and $t \oplus r = 1$. After that, Eve performs $U_F$ on the particles sent back to TP. For Eve not being detectable in Step 5, $U_F$ should satisfy

$$U_F(|t\rangle_5|r\rangle_6|\xi\rangle) = |t\rangle_5|r\rangle_6|\delta_{tr}\rangle \tag{41}$$

for $t, r \in \{0,1\}$ and $t \oplus r = 1$, which means that $U_F$ cannot alter the state of particles from Alice and Bob. Otherwise, Eve is discovered with a non-zero probability.

(ii) Consider the case that Alice chooses to MEASURE and Bob chooses to REFLECT. As a result, the state of the composite system is collapsed into $|0\rangle_5|1\rangle_6|\xi\rangle_E$ when Alice's measurement result is $|0\rangle_5$ or $|1\rangle_5|0\rangle_6|\xi\rangle_E$ when Alice's measurement result is $|1\rangle_5$.

Assume that Alice's measurement result is $|0\rangle_5$. After Eve imposes $U_F$ on the particles sent back to TP, due to Eq.(41), the state of the composite system is evolved into

$$U_F(|0\rangle_5|1\rangle_6|\xi\rangle) = |0\rangle_5|1\rangle_6|\delta_{01}\rangle. \tag{42}$$

For Eve not being detectable in Step 5, the probability that TP's measurement result on the particle reflected by Bob is $|0\rangle_6$ should be 0. Apparently, this point is automatically satisfied.

On the other hand, assume that Alice's measurement result is $|1\rangle_5$. After Eve imposes $U_F$ on the particles sent back to TP, due to Eq.(41), the state of the composite system is evolved into

$$U_F(|1\rangle_5|0\rangle_6|\xi\rangle) = |1\rangle_5|0\rangle_6|\delta_{10}\rangle. \tag{43}$$

For Eve not being detectable in Step 5, the probability that TP's measurement result on the particle reflected by Bob is $|1\rangle_6$ should be 0. Apparently, this point is automatically satisfied.

(iii) Consider the case that Alice chooses to REFLECT and Bob chooses to MEASURE. As a

result, the state of the composite system is collapsed into $|1\rangle_5 |0\rangle_6 |\xi\rangle_E$ when Bob's measurement result is $|0\rangle_6$ or $|0\rangle_5 |1\rangle_6 |\xi\rangle_E$ when Bob's measurement result is $|1\rangle_6$.

Assume that Bob's measurement result is $|0\rangle_6$. After Eve imposes $U_F$ on the particles sent back to TP, due to Eq.(41), the state of the composite system is evolved into Eq.(43). For Eve not being detectable in Step 5, the probability that TP's measurement result on the particle reflected by Alice is $|0\rangle_5$ should be 0. Apparently, this point is automatically satisfied.

On the other hand, assume that Bob's measurement result is $|1\rangle_6$. After Eve imposes $U_F$ on the particles sent back to TP, due to Eq.(41), the state of the composite system is evolved into Eq.(42). For Eve not being detectable in Step 5, the probability that TP's measurement result on the particle reflected by Alice is $|1\rangle_5$ should be 0. Apparently, this point is automatically satisfied.

(iv) Consider the case that both Alice and Bob choose to REFLECT. As a result, the state of the composite system is $|\psi^+\rangle_{56} |\xi\rangle_E$. After Eve imposes $U_F$ on the particles sent back to TP, due to Eq.(41), the state of the composite system is evolved into

$$U_F \left(|\psi^+\rangle_{56} |\xi\rangle\right) = \frac{1}{\sqrt{2}} \left(|0\rangle_5 |1\rangle_6 |\delta_{01}\rangle + |1\rangle_5 |0\rangle_6 |\delta_{10}\rangle\right)$$
$$= \frac{1}{2} \left(|\psi^+\rangle_{56} + |\psi^-\rangle_{56}\right) |\delta_{01}\rangle + \frac{1}{2} \left(|\psi^+\rangle_{56} - |\psi^-\rangle_{56}\right) |\delta_{10}\rangle$$
$$= \frac{1}{2} |\psi^+\rangle_{56} \left(|\delta_{01}\rangle + |\delta_{10}\rangle\right) + \frac{1}{2} |\psi^-\rangle_{56} \left(|\delta_{01}\rangle - |\delta_{10}\rangle\right). \tag{44}$$

For Eve not being detectable in Step 5, the probability that TP's measurement result on the particles reflected by Alice and Bob is $|\psi^+\rangle_{56}$ should be 1. Hence, it can be obtained from Eq.(44) that

$$|\delta_{01}\rangle = |\delta_{10}\rangle = |\delta\rangle. \tag{45}$$

Inserting Eq.(45) into Eq.(44) generates

$$U_F \left(|\psi^+\rangle_{56} |\xi\rangle\right) = |\psi^+\rangle_{56} |\delta\rangle. \tag{46}$$

It can be concluded from the above analysis: considering the entangle-measure attack from Eve that she only performs $U_F$ on the particles of $T_5$ and $T_6$ sent back to TP, for Eve not inducing an error in Step 5, the final state of Eve's probe should be independent of not only Alice and Bob's operations but also their measurement results. As a result, Eve gets no information on the bits of $S_5$ and $S_6$.

(4) The Trojan horse attack

In accordance with Refs.[57,58], the invisible photon eavesdropping attack and the delay-photon Trojan horse attack from Eve on the particles of $T_5$ during the round trip between TP and Alice can be overcome by Alice with a wavelength filter and a photon number splitter, respectively. The same methods can be used to prevent the invisible photon eavesdropping attack and the delay-photon Trojan horse attack from Eve on the particles of $T_6$ during the round trip between TP and Bob.

**4.2 Participant attack**

With respect to the participant attack suggested by Gao *et al.* [59], we need to consider the participant attack from Alice or Bob and that from the semi-honest TP.

(1) The participant attack from Alice or Bob

In the proposed protocol, Alice plays the same role to Bob. Without loss of generality, we only consider the case that Bob, who is supposed to have complete quantum abilities, is dishonest.

In the proposed protocol, Bob naturally knows $S_3$ and $S_6$. Moreover, according to the entanglement correlation of two qubits within one Bell state, Bob can deduce $S_4$ and $S_5$ from $S_3$ and $S_6$, respectively. As $M_A$ is encrypted with $S_1$, $S_5$ and $K_{AB}$, in order to deduce $M_A$ from $R_A$, Bob should further know $S_1$. However, Bob cannot get $S_1$ by cooperating with TP who knows $S_2$. As a result, Bob has to try his best to get $S_1$ by launching some active attacks on the particles of $T_1$, such as the intercept-resend attack, the measure-resend attack, the entangle-measure attack, the Trojan horse attack, *et al.* However, he will be inevitably discovered as an external eavesdropper, since Alice's operations are random to him, after similar deductions to those of Case 1 in Section 4.1. In conclusion, Bob has no chance to obtain $M_A$.

(2) The participant attack from TP

TP may try to obtain $M_A$ and $M_B$ by using passive attacks. In other words, TP may try to reveal $M_A$ and $M_B$ from the classical information she collected during the implementation of the protocol. In the proposed protocol, TP can know $S_1, S_2, S_3, S_4, S_5, S_6, R_A, R_B$ and the comparison result of $M_A$ and $M_B$. However, TP still cannot deduce $M_A$ from $R_A$ and $M_B$ from $R_B$, because she cannot obtain $K_{AB}$ shared by Alice and Bob beforehand which is used to encrypt $M_A$ and $M_B$.

In the proposed protocol, TP is assumed to be semi-honest. As a result, she may try to perform possible active attacks to get $M_A$ and $M_B$ but cannot collude with anyone else. For example, TP may prepare fake quantum states, such as product states or nonmaximal entangled states, instead of Bell states in Step 1. The proposed protocol lacks the process of checking whether TP has prepared the genuine Bell states in Step 1 or not, hence, Alice and Bob may be unknown about TP's cheating behavior. Actually, no matter what kinds of attack TP launches, the best result for TP is that she can obtain $S_1, S_2, S_3, S_4, S_5, S_6, R_A, R_B$ and the comparison result of $M_A$ and $M_B$. However, TP still has no opportunity to decode out $M_A$ from $R_A$ and $M_B$ from $R_B$, due to lack of $K_{AB}$.

## 5  Discussions and conclusions

In this part, we compare the proposed protocol with the existing SQPC protocols based on Bell states in Refs.[39,43,44,48-50,53]. The detailed comparison results are shown in Table 4. Here, the qubit efficiency is defined as [60] $\eta = \frac{b}{q+c}$, where $b$, $q$ and $c$ represent the numbers of compared private bits, consumed qubits and classical bits involved in classical communication, respectively. Note that the classical resources required for eavesdropping detection are not considered here.

In our protocol, Alice and Bob successfully compare their respective $n$ private bits, so it has $b = n$. TP prepares $N = 16n$ initial Bell states and distributes $T_1$ and $T_5$ to Alice and $T_3$ and $T_6$ to Bob. Then, when Alice chooses to MEASURE the received qubits in $T_1$ and $T_5$, she prepares $2n$ and $4n$ new qubits, respectively. When Bob chooses to MEASURE the received qubits in $T_3$ and $T_6$, he also prepares $2n$ and $4n$ new qubits, respectively. In addition, this protocol employs the SQKD protocol in Ref.[31] to generate the $n$-bit pre-shared key $K_{AB}$, which consumes $24n$ qubits. As a result, it has $q = 16n \times 2 + 2n \times 2 + 4n \times 2 + 24n = 68n$. Alice sends $R_A$ to TP, while Bob sends $R_B$ to TP. Hence, it has $c = 2n$. Therefore, the qubit efficiency of our protocol is $\eta = \frac{n}{68n+2n} = \frac{1}{70}$.

In the protocol of Ref.[39], Alice and Bob successfully compare their respective $n$ bits of hash values, so it has $b = n$. Server prepares $4n$ initial Bell states and distributes their first and second particles to Alice and Bob, respectively. Then, Alice and Bob save the measurement results or reflect the received qubits back after reordering them. As a result, it has $q = 4n \times 2 = 8n$. Alice sends $R_A$ to Bob, while Bob sends $R_B$ to Alice. Hence, it has $c = 2n$. Therefore, the qubit efficiency of the protocol of Ref.[39] is $\eta = \frac{n}{8n+2n} = \frac{1}{10}$. Note that after Alice and Bob publish $R_A$ and $R_B$, respectively, Server can know the comparison result by computing $R_A \oplus R_B$.

In the protocol of Ref.[43], Alice and Bob successfully compare their respective $n$ private bits, so it has $b = n$. TP prepares two sequences of Bell states, each of whose length is $16n$, and distributes the first particles of two sequences to Alice and Bob, respectively. Then, when Alice and Bob choose to measure the received qubits, they replace the measurement results with the freshly prepared qubits. As a result, it has $q = 16n \times 2 \times 2 + 8n \times 2 = 80n$. Alice sends $R_A$ to TP, while Bob sends $R_B$ to TP. Hence, it has $c = 2n$. Therefore, the qubit efficiency of the protocol of Ref.[43] is $\eta = \frac{n}{80n+2n} = \frac{1}{82}$.

In the protocol of Ref.[44], Alice and Bob successfully compare their respective $n$ private bits,

so it has $b = n$. TP prepares $8n$ initial Bell states and distributes their first and second particles to Alice and Bob, respectively. Then, when Alice and Bob choose to MEASURE the received qubits, they replace the measurement results with the freshly prepared qubits. In addition, the protocol employs the SQKD protocol in Ref.[31] to generate the $n$-bit pre-shared key $K_{AB}$, which consumes $24n$ qubits, and the SQKA protocol in Ref.[41] to generate the $n$-bit pre-shared keys, $K_{AT}$ and $K_{BT}$, which consumes $10n$ qubits in total. As a result, it has $q = 8n \times 2 + 4n \times 2 + 24n + 10n = 58n$. Alice sends $C_A$ to TP, while Bob sends $C_B$ to TP. Hence, it has $c = 2n$. Therefore, the qubit efficiency of the protocol of Ref.[44] is $\eta = \frac{n}{58n + 2n} = \frac{1}{60}$.

In the second protocol of Ref.[48], Alice and Bob successfully compare their respective $n$ private bits, so it has $b = n$. TP prepares $2n$ initial Bell states and distributes their first and second particles to Alice and Bob, respectively. Then, when Alice and Bob choose to flip the received qubits, they replace the measurement results with the freshly prepared opposite qubits. In addition, the protocol employs the SQKD protocol in Ref.[31] to generate the $n$-bit pre-shared key $K$, which consumes $24n$ qubits. As a result, it has $q = 2n \times 2 + n \times 2 + 24n = 30n$. Alice sends $M_A$ to TP, while Bob sends $M_B$ to TP. Hence, it has $c = 2n$. Therefore, the qubit efficiency of the second protocol of Ref.[48] is $\eta = \frac{n}{30n + 2n} = \frac{1}{32}$.

In the protocol of Ref.[49], Alice and Bob successfully compare their respective $n$ private bits, so it has $b = n$. TP prepares $2n$ initial Bell states and distributes their first and second particles to Alice and Bob, respectively. Then, when Alice and Bob choose to SIFT, they generate fresh qubits and send them to TP. In addition, the protocol uses the SQKD protocol in Ref.[29] to generate two $n$-bit pre-shared keys, $K_{TA}$ and $K_{TB}$, which consumes $16n$ qubits in total, and the SQKD protocol in Ref.[31] to generate the $n$-bit pre-shared key $K$, which consumes $24n$ qubits. As a result, it has $q = 2n \times 2 + n \times 2 + 16n + 24n = 46n$. Alice publishes $R_A$ to TP, while Bob publishes $R_B$ to TP. Hence, it has $c = 2n$. Therefore, the qubit efficiency of the protocol of Ref.[49] is $\eta = \frac{n}{46n + 2n} = \frac{1}{48}$.

In the protocol of Ref.[50], Alice and Bob successfully compare their respective $n$ private bits, so it has $b = n$. TP prepares $3n$ initial Bell states and distributes their first and second particles to Alice and Bob, respectively. Then, when Alice and Bob choose to SIFT, they generate fresh qubits and send them to TP. When Alice and Bob choose to DETECT, they generate fresh trap qubits and send them to TP. In addition, the protocol uses the SQKD protocol in Ref.[31] to generate the $n$-bit pre-shared key $K$, which consumes $24n$ qubits. As a result, it has $q = 3n \times 2 + n \times 2 + n \times 2 + 24n = 34n$. Alice publishes $R_A \oplus R_A^{'}$ to TP, while Bob publishes $R_B \oplus R_B^{'}$ to TP. Hence, it has $c = 2n$. Therefore, the qubit efficiency of the protocol of Ref.[50] is $\eta = \frac{n}{34n + 2n} = \frac{1}{36}$.

In the protocol of Ref.[53], Alice and Bob successfully compare their respective $n$ private bits,

so it has $b=n$. TP prepares $2n$ initial Bell states and distributes their first and second particles to Alice and Bob, respectively. Then, when Alice and Bob choose to MEASURE the received qubits, they generate fresh qubits and send them to TP. In addition, this protocol utilizes the SQKD protocol in Ref.[31] to generate the $2n$-bit pre-shared key $K_{AB}$, which consumes $48n$ qubits. As a result, it has $q=2n\times 2+n\times 2+48n=54n$. Both Alice and Bob publish $K_{AB}$ to TP. Hence, it has $c=4n$. Therefore, the qubit efficiency of the protocol of Ref.[53] is $\eta=\frac{n}{54n+4n}=\frac{1}{58}$.

In addition, Ref.[43] needs quantum entanglement swapping technology, which may be difficult to implement in practice. Fortunately, our protocol doesn't need quantum entanglement swapping technology. All of the SQPC protocols in Refs.[39,44,48] need to perform the reordering operations via different delay lines which require complicated quantum circuits in practice. Fortunately, our protocol doesn't have this requirement. Moreover, each of the SQPC protocols in Refs.[43,44,48-50,53] needs to prepare four kinds of Bell states as initial quantum resource. Fortunately, our protocol only needs to generate one kind of Bell states. It is naturally that preparing the same kind of Bell states repeatedly is easier in practice than generating different types of Bell states. To sum up, our protocol is easier to implement in practice than the SQPC protocols in Refs.[39,43,44,48-50,53].

In addition, here we can further take into account that TP may announce the fake comparison result in Step 7, which cannot be detected by Alice and Bob. In order to overcome this cheating behavior from TP, the following extra supplement should be added to the above proposed protocol: (1) in Step 6, while Alice and Bob sends $R_A$ and $R_B$ to TP, respectively, they also send $R_A$ and $R_B$ to the second semi-honest third party, TP2; (2) TP2 shares a secret key $K_T$ with TP via a secure QKD scheme, where $K_T=\{K_T^1,K_T^2,\cdots,K_T^n\}$; (3) in Step 7, TP computes $R_{TP}^j=K_T^j\oplus S_2^j\oplus S_4^j$ and sends $R_{TP}$ to TP2, then TP2 computes $R_{T2}^j=R_A^j\oplus R_B^j\oplus R_{TP}^j\oplus K_T^j$ after receiving $R_{TP}$, where $R_{TP}=\{R_{TP}^1,R_{TP}^2,\cdots,R_{TP}^n\}$ and $j=1,2,\cdots,n$; once $R_{T2}^j=0$ is found, TP2 tells Alice and Bob that $M_A$ is not identical to $M_B$, otherwise, TP2 tells Alice and Bob that $M_A$ is identical to $M_B$. Apparently, because TP and TP2 cannot conspire, their published comparison results are mutually independent. Alice and Bob can judge whether TP and TP2 have announced fake comparison results or not by comparing the consistency of their published comparison results.

In conclusion, in this paper, a novel SQPC protocol based on single kind of Bell states is proposed, which utilizes the entanglement correlation of Bell states to skillfully compare the equality of private inputs from two classical users. Our protocol can resist a variety of outside and participant attacks. Our protocol only employs one kind of Bell states as initial quantum resource. Moreover, it needs none of unitary operations, quantum entanglement swapping or the reordering operations. Therefore, compared with most of the existing SQPC protocols with Bell states, our protocol has better practical implementation feasibility.

Table 4  Comparison results of our SQPC protocol and the previous SQPC protocols with Bell states

|  | The protocol of Ref.[39] | The protocol of Ref.[43] | The protocol of Ref.[44] | The second protocol of Ref.[48] | The protocol of Ref.[49] | The protocol of Ref.[50] | The protocol of Ref.[53] | Our protocol |
|---|---|---|---|---|---|---|---|---|
| Feature | Randomization-based | Measure-resend | Measure-resend | Measure-Randomization-resend | Discard-resend | Measure-discard-resend | Measure-resend | Measure-resend |
| Types of Bell states | Single | Four | Four | Four | Four | Four | Four | Single |
| Usage of SQKD or SQKA | No | No | Yes | Yes | Yes | Yes | Yes | Yes |
| Usage of quantum entanglement swapping | No | Yes | No | No | No | No | No | No |
| Usage of unitary operations | No | No | No | No | No | No | No | No |
| Usage of delay lines | Yes | No | Yes | Yes | No | No | No | No |
| TP's knowledge about the comparison result | Yes | Yes | Yes | Yes | Yes | Yes | Yes | Yes |
| Qubit efficiency | $\frac{1}{10}$ | $\frac{1}{82}$ | $\frac{1}{60}$ | $\frac{1}{32}$ | $\frac{1}{48}$ | $\frac{1}{36}$ | $\frac{1}{58}$ | $\frac{1}{70}$ |

**Funding:** the National Natural Science Foundation of China (Grant No.62071430 and No.61871347) and the Fundamental Research Funds for the Provincial Universities of Zhejiang (Grant No.JRK21002);

**Acknowledgements:** The authors would like to thank the anonymous reviewers for their valuable comments that help enhancing the quality of this paper.